Journalists, Emotions, and the Introduction of Generative AI Chatbots:

A Large-Scale Analysis of Tweets Before and After the Launch of ChatGPT

**Seth C. Lewis, Ph.D.**
Professor and Shirley Papé Chair in Emerging Media
School of Journalism and Communication
University of Oregon (USA)
sclewis@uoregon.edu

**David M. Markowitz, Ph.D.**
Associate Professor
Department of Communication
Michigan State University (USA)
dmm@msu.edu

**Ben Bunquin**
Doctoral Candidate
School of Journalism and Communication
University of Oregon (USA)
jbunquin@uoregon.edu

**Abstract**

As part of a broader look at the impact of generative AI, this study investigated the emotional responses of journalists to the release of ChatGPT at the time of its launch. By analyzing nearly 1 million Tweets from journalists at major U.S. news outlets, we tracked changes in emotional tone and sentiment before and after the introduction of ChatGPT in November 2022. Using various computational and natural language processing techniques to measure emotional shifts in response to ChatGPT's release, we found an increase in positive emotion and a more favorable tone post-launch, suggesting initial optimism toward AI's potential. This research underscores the pivotal role of journalists as interpreters of technological innovation and disruption, highlighting how their emotional reactions may shape public narratives around emerging technologies. The study contributes to understanding the intersection of journalism, emotion, and AI, offering insights into the broader societal impact of generative AI tools.

**Journalists, Emotions, and the Introduction of Generative AI Chatbots:**

**A Large-Scale Analysis of Tweets Before and After the Launch of ChatGPT**

With the introduction of any new technology, particularly one that appears to be disruptive or transformative in nature, there is a tendency for people to emphasize one of two competing hypotheses about the technology and its impact: dystopian vs. utopian (Nowland et al., 2018; Valkenburg & Peter, 2007). At the risk of oversimplifying, we might say that one camp highlights, often in terms of anxiety, fear, and worry, the risks and potentially negative consequences of the technology. Meanwhile, another camp focuses more on the perceived benefits for individuals and society, often emphasizing hope, optimism, awe, and amazement.

Following this pattern, the launch and then rapid mainstream adoption of chatbots like ChatGPT has engendered all manner of hope as well as hysteria, astonishment as well as anxiety—with various detractors and supporters alike emphasizing, on the one hand, the potentially deleterious outcomes of large language models (LLMs) for jobs, education, and notions of truth and reality, and, on the other, the expansive possibilities for creativity, education, economic productivity, and more (e.g., see Mollick, 2024; Vrabič Dežman, 2024). While not the only LLM chatbot available, ChatGPT quickly became the most visible and attention-grabbing after its public release in late 2022. In less than a week after its launch, ChatGPT reached 1 million users, and by January 2023 it had an estimated 100 million active users, making it one of the fastest-adopted consumer applications in history.

If artificial intelligence in the broad sense refers to the use of machines to perform tasks typically requiring human intelligence, often by learning from experience, parsing natural language, recognizing patterns, and solving problems (Boden, 2018; Broussard, 2018; Mitchell, 2019), then ChatGPT and tools like it represent a specialized subset of AI called *generative AI*, so named because they involve the generation of new content—such as text, images, audio, video, or code—at unprecedented speed and scale. LLMs such as Claude 3.5 Sonnet from Anthropic or GPT-4 from OpenAI, for example, are advanced generative AI technologies that have been trained on huge volumes of text data that then allow users—such as people prompting ChatGPT—to generate human-like text on command. These developments in generative AI, in particular, have come to be seen as revolutionary, whether for good or ill, across many industries and domains. Many business executives talk about the technology's potential to transform hiring (Kelly, 2023) and customer service (Das et al., 2023), even as the same tools and capabilities have elicited fears of replacement among workers, with some estimates of 1 in 5 jobs (or more) being affected by some form of displacement in the near term (Hatzius et al., 2023).

Ultimately, the actual impact of generative AI on jobs, industries, and daily life broadly may depend in large part on the stories that people tell themselves about disruptive technology, and such stories may reflect the tone—pro or con, hype or hysteria—that may be set by elite influencers in society (cf. Yin, 1999). Against this backdrop, this paper asks: Who ultimately shapes the formation and perpetuation of these narratives—and, therefore, why do they matter? We address this question by investigating how tastemakers like journalists responded to the mainstreaming of generative AI as indicated by emotion and tone expressed on social media.

Journalists offer a particularly useful case for several reasons. For many decades, research has demonstrated the power of the news media to set the agenda for public discussion (McCombs & Valenzuela, 2020), though journalists can be influential not only in drawing attention to certain topics over others but also in how they emphasize various qualities and characteristics of the people, issues, and ideas being discussed (Entman et al., 2009). Indeed, well-established is the distinct role of journalists as "sense-makers" for society (Singer, 1998).

We see this in the way journalists help shape the terms of debate, establish narratives that persist over time (e.g., the so-called "first draft of history"), and offer framing devices that altogether matter for influencing public opinion formation (Lecheler & de Vreese, 2012)—not in an entirely deterministic way, but nevertheless in a way that has been shown to demonstrably matter (Entman et al., 2009). Even in the increasingly fragmented media landscape of the 21st century, in which news media must fight to claim a sufficient share of audience attention, there remains substantial evidence that journalists still matter as tastemakers for society (Langer & Gruber, 2021; McCombs & Valenzuela, 2020; see also Schudson, 2018).

Importantly, the narratives constructed by journalists often function as "fictional expectations" (Beckert, 2016) used for decision-making: that is, the expected values or rewards can orient decision-making toward the attainment (or avoidance) of certain imagined futures. Thus, the prevailing narratives, to which journalists and the media at large serve as key contributors, not only impact current conceptions of these technologies but also shape ideas about and expectations from AI that are used as reference points in shaping future outcomes. These narratives, however, are not purely rational or fully formed; rather, the emotional states of journalists themselves may serve as crucial elements in forming the imaginations that take shape and become collectively adopted in society. As such, the individual and collective emotions expressed by tastemakers such as journalists—whether infused with optimism, skepticism, or alarm—contribute to the types of futures that are ultimately envisioned and conveyed through news narratives. Put another way, journalists are more than just reporters—they are instructors on how others should think and feel about the information being reported.

One way to tap into the imagined futures of journalists that has yet to be explored, particularly at any large scale, is to study the emotions of journalists. Emotion, or the discrete feeling states that represent reactions towards events (Barrett, 2017), can serve as a valuable marker for signaling our hopes and desires as well as our fears and anxieties—thereby offering a lens into how people feel about the future in a way that helps to render that future actualizable. Studying emotions, as customary in disciplines like communication science (Nabi, 2010) and psychology (Cacioppo & Gardner, 1999), thus provides deeper insights into the motivational forces driving journalistic perspectives and the narratives they create, which have consequences on public perception and discourse about the future.

This study examines markers of journalists' emotions at scale, and it does so at a particularly apt moment: the periods before and after the introduction of ChatGPT. Our goal is to use this exploration of emotions to investigate the competing (and pervasive) hypotheses about the increasing role and influence of generative AI: Is the doom-and-gloom hypothesis of dystopian concern more prominent, perhaps because journalists fear how such tools, given their ability to interpret and generate language, might displace their work through automation? Or, is the optimistic hypothesis more prominent because journalists, as some evidence indicates thus far, see opportunities in generative AI to free them up by taking on mundane tasks that could be automated? An examination of the emotions that are evident in journalists' expressions online— and especially of how such emotions are manifest in the case of journalists who are most preoccupied with these issues of ChatGPT and AI—can therefore offer a window into how publics broadly may come to form narratives about the imagined futures associated with these technologies. In effect: If journalists set the tone for how people may come to imagine these technologies and what they expect from them, then what kind of tone, positive or negative, is being established? And what might that reveal about the role of emotions in making judgments about the imagined futures of new technologies?

## Literature Review and Contextual Background

### AI and Imagined Futures

Since November 30, 2022, when the research firm and tech company OpenAI launched ChatGPT to much public fanfare, social media conversations about artificial intelligence have been dominated by talk of ChatGPT's role and influence (Maslej et al., 2023), intensifying broader public discussion about AI generally and generative AI in particular. The potential loss of human jobs due to AI stands out as a top concern among many Americans (Maslej et al., 2023), and the implications of AI for job security have become more pronounced, particularly given the demonstrated efficiency of ChatGPT and other generative AI products. This narrative is also evident in the media, where numerous articles have been published discussing jobs that could be replaced by AI after the launch of ChatGPT (e.g., Mitchell, 2023; Stahl, 2023).

News accounts such as these constitute part of how societies imagine the future in relation to technological developments. These imaginations, born out of people's expectations of technologies, inform current practices and systems, which are enacted to attain or avoid such futures (Jasanoff & Kim, 2009). In other words, expectations about the future mobilize changes within the present, and tastemakers such as journalists serve as intermediaries not only in shaping these expectations, but also in producing new ones, as they possess significant influence over the narratives surrounding new, disruptive technologies (Brennen et al., 2022).

Long before the introduction of ChatGPT, future imaginations around AI had been circulating in society, and they fall into two parallel narratives of hope and fear (Cave & Dihal, 2019). The promise of longevity and immortality is one of these hopeful narratives, with news media reproducing this imaginary of an AI expected to make significant contributions to healthcare and medicine (Cave & Dihal, 2019; Wang et al., 2023). AI is also expected to make living conditions easier (Hancock et al., 2020), as everyday routine tasks are automated, essentially making AI serve as our assistants (Cave & Dihal, 2019). In this vision of AI, as humans are freed from doing mundane activities, AI helps people live more meaningful lives (Hautala & Heino, 2023). Additionally, AI is expected to be a tool that can be effectively harnessed to drive innovations, augment economic and social capacity of humans, provide more opportunities for employment (Hautala & Heino, 2023; Wang et al., 2023), promote efficient systems and sustainability, and help nations advance in governance and labor productivity (Bareis & Katzenbach, 2022).

Inherent in these hopeful imaginaries are instabilities that can lead to dystopian outcomes, represented as narratives of fear. Opposite longevity and immortality is the possibility of inhumanity or the risk of losing human identity. Meanwhile, the ease of work afforded through AI is met with fears of obsolescence from work and being replaced by technologies. The uprising of machines is also anticipated due to the technology (Cave and Dihal, 2019), viewing the possibility of artificial intelligence taking over humanity and controlling humans "in a matrix-like world" (Hautala & Heino, 2023, p. 5). Others imagine AI as a technology that people will use to take advantage of others and widen social inequalities, with corporations owning AI development as well as data, and humans becoming unemployed and further marginalized (Hautala & Heino, 2023). The rise of AI is also viewed as the end of humanity, and this is reflected in a 22-word statement signed by many scientists, public figures, and industry leaders which states that "Mitigating the risk of extinction from AI should be a global priority alongside other societal-scale risks such as pandemics and nuclear war" (*Statement on AI Risk*, 2023).

In both hopeful and fearful narratives, expectations about AI—and media or technology in general—are fueled by affect and emotion, but it is more pronounced in the dystopian rather than the utopian views of the technology (Nowland et al., 2018). Inaccurate portrayals and anthropomorphization of AI, conflating human autonomy with computer autonomy, and socio-technical blindness, or the failure to recognize how technologies always work in tandem with humans and institutions, are believed to contribute to AI anxiety, or the "fear and trepidation being expressed about out-of-control AI" (Johnson & Verdicchio, 2017, p. 2267). The role of journalists as tastemakers is further highlighted as they make these anxieties evident through their outputs (Brennen et al., 2022), and contribute further to the apprehensions and fear narratives revolving around seemingly uncontrollable technologies (Sartori & Bocca, 2023). Understanding the role of emotions on public perceptions of AI, and the significant role that journalists play in shaping these emotional narratives, underscores the importance of studying the intersection of emotions and journalism in the context of artificial intelligence.

**The Import of Affect and Emotion: A Psychology of Language Perspective**

Classic social scientific theories of emotion suggest that there are six dominant emotional expressions (sadness, happiness, fear, anger, surprise, disgust) (Ekman et al., 1987), which represent feelings we have to socially constructed events (Barrett, 2017). The "goodness" or "badness" of something, also known as affect (Slovic et al., 2004), is related to emotion, but it is an independent concept. For the purposes of our work, we use *emotion* to mean discrete emotions from Ekman, Barrett, and others, while *affect* is a broader understanding of the positive or negative attributes of an object that includes (but is not limited to) discrete emotions.

The scientific study of emotion and affect is over a century old (James, 1884), with keen interests in how they are constructed (Barrett, 2017), regulated (Gross, 1998), spread (Kramer et al., 2014), how they guide decision-making (Nabi, 2003; Peters et al., 2006), and also associate with a range of social and psychological processes (Barrett et al., 2007; Cacioppo & Gardner, 1999). In many cases, emotion and affect are studied as antecedents or consequences of key psychological events. For example, people study emotions to understand how they guide health choices (Peters et al., 2006) and how emotion can be the result of trauma or upheavals (Galea et al., 2020; Markowitz, 2022; Seraj et al., 2021). We adopt a similar approach and attempt to understand emotion and affect as a consequence of technological innovation. The feelings people have about technology—reflecting either utopian or dystopian viewpoints—form a narrative about how the public should also think and feel about technology. This agenda-setting, trickle-down perspective is essential to assess at this paradigm-shifting moment with the rise of chatbots, large language models, and generative AI. We take an innovative approach to this research agenda by tracking emotion in language patterns, at scale, and on social media.

The idea that language patterns can reveal important social and psychological information about people originates from a research tradition that suggests words serve as markers of attention and focus (Boyd & Schwartz, 2021; Boyd & Markowitz, 2024; Pennebaker, 2011). For example, word patterns have been linked to personality traits and individual differences (Ireland & Mehl, 2014; Newman et al., 2008; Pennebaker & King, 1999), well-being (Jaidka et al., 2020; Rude et al., 2004; Stirman & Pennebaker, 2001), societal discourse shifts in institutions (Jordan et al., 2019), and the main focus of the present article, emotion and affect (Doyle et al., 2021; Sauter, 2018; Vine et al., 2020). From this perspective, words indicate where or what people are attending to and where or what they are not attending to. A person who suggests "I hate technology" is focusing on a negative emotional state (e.g., using the word *hate*), but scholars in this tradition would not go so far as to suggest they *felt* negative. This words-as-attention model to understanding people through language is pervasive in the social sciences (Boyd & Schwartz, 2021; Boyd & Markowitz, 2024), with studies using words as a lens in humans, their psychological states, and internal processing of events.

Drawing on this approach, a range of studies suggest emotion can serve as an integral marker of social and societal-level change and can be evaluated linguistically at scale. For example, prior work tracked the linguistic patterns of bloggers after the 9/11 terrorist attacks and observed a sharp increase in emotion immediately after the event, returning to baseline weeks later (Cohn et al., 2004). Other work has found that, compared to before the COVID-19 pandemic, academics wrote with a more emotional focus in their journal articles during the COVID-19 pandemic (Markowitz, 2022). Specifically, those who wrote about COVID-19 during the pandemic had the greatest rate of negative emotion compared to those writing before the pandemic or during the pandemic but about other topics. Altogether, dozens of studies suggest that by tracking how people represent emotion in language, we can understand important social and psychological information about communicators and how they perceive the world. We are among the first to apply this large-scale, computational approach to the understanding of how journalists write about new technology such as large language models and chatbots.

**Journalists, Disruptive Technologies, and Twitter (now X)**

Language patterns can unveil how journalists make sense of disruptive technologies, offering insights about their perceptions and expectations of technology plus how those attitudes could impact audiences. Previous research on journalists and their responses to disruptive technologies can provide an initial understanding of their expectations about new technology.

Journalists tend to hold varying expectations about disruptive technologies. As the internet became widely used in the early 2000s, and as new multimedia tools for reporting emerged alongside the development of the World Wide Web, such technologies were met with optimism by journalists, who imagined how new media might improve efficiency in collecting information and connecting with sources (Chadha & Wells, 2016; O'Sullivan & Heinonen, 2008; Pont-Sorribes et al., 2013). New technologies indeed have aided reporting: databases are used to help journalists anticipate and predict newsworthy events (Linden, 2017); automated journalism can convert structured data about topics like sports, weather, and finance into narrative articles with little human involvement (Diakopoulos, 2019); and social media platforms amplify news distribution and improve opportunities for audience interaction as well as sourcing (Lewis & Molyneux, 2018).

Despite these efficiency expectations, journalists also have expressed ample concern about how emerging technologies might degrade their profession. Journalists for many years have worried about a profusion of low-quality information online (O'Sullivan & Heinonen, 2008), do-it-yourself approaches to reporting that undermine the value of professionals (Posseti, 2009), and the likelihood that "robot journalists" (i.e., software) would produce stories with algorithmically introduced errors (Kim & Kim, 2018), highlighting the conflict between immediacy afforded by new technologies and accuracy prioritized in journalism (Linden, 2017). More existentially, journalists have been alarmed that the growing use of automation in journalism could threaten their future employment, or at least make journalists become overly reliant on machines (Kim & Kim, 2018; Linden, 2017; Moran & Shaikh, 2022).

As disruptive technologies become integrated in newsrooms, journalists also anticipate an increase in their obligations—a "hamster wheel" of evermore work to be done (Usher, 2016; cf. Bélair-Gagnon et al., 2022, 2024). Some reporters resist by adhering to a "principle of continuity" (O'Sullivan & Heinonen, 2008, p. 367), sticking to conventions and old habits that have long worked for them (see also Powers & Vera-Zambrano, 2019). Other journalists recognize that new technologies could help advance their careers, but that such potential might only be realized if they have the training and resources to achieve proficiency (Powers & Vera-Zambrano, 2019), which is not easily accomplished in an era of cutbacks for many newsroom staffs (Ferrucci & Perreault, 2021).

Of all the technologies to emerge in recent decades, perhaps none has been more widely adopted by journalists than social media, used as an integral tool in news production and distribution. Platforms like Twitter, Facebook, and TikTok have played crucial roles in breaking news stories and disseminating them, and media organizations have strategically engaged the networked architectures and algorithmic design of such platforms to reach audiences where they are (Hermida, 2018; Lewis & Molyneux, 2018). Twitter, in particular, has emerged as journalists' preferred social media platform, especially in much of the English-speaking West, because its continuous "stream of news, comments and analysis" makes it a handy, always-on source for news and information (Hermida, 2018, p. 4). Journalists have played an outsized role on Twitter, serving as a "real-time interpretive community" (Araiza et al., 2016, p. 310), blending personal and professional identities in the way they share and discuss the news online, and relying on each other and other users to make sense of events in real-time (see discussion in Mellado & Hermida, 2021). Twitter has also been used extensively by journalists as a de facto public sphere, as they draw on the platform to collect a "modern version of person-on-the-street interviews" (Lewis & Molyneux, 2018, p. 16). Indeed, journalists, in the U.S. especially, have become so reliant on Twitter that some have questioned what this dependence has done to cloud journalists' sense of news judgment (McGregor & Molyneux, 2020).

In 2022, Elon Musk bought Twitter for $44 billion. At the time, "journalists around the world looked on in alarm," as Musk, who had a fractious relationship with journalists, threatened to dismantle an existing blue-check verification system that favored news media (Gotfredsen,

2023). Many journalists (as well as academics; see Braun, 2024) reacted by threatening to leave the platform for other online communities. And while some did, an analysis in early 2023 of some 4,000 journalists from 19 U.S. news outlets found that only a small fraction of journalists actually deactivated their accounts—and, on average overall, journalists were tweeting about 3% less after Musk's takeover of the platform (Gotfredsen, 2023). In mid-2023, Musk renamed the service X, but colloquially many people still refer to the platform as Twitter.

Journalists continue to use Twitter (now X) not only for professional practice, but also for their personal agendas. By sharing behind-the-scenes information with their followers, they build stronger, more personal relationships with their audiences and may enhance their reach and reputation (Molyneux, 2019; Mellado & Alfaro, 2020; Mellado & Hermida, 2021). Because of their engagement on the platform, some journalists have even become political influencers, tailoring content to what their audiences want based on what they glean as relevant through interactions with their followers (Peres-Neto, 2022). Despite newsroom guidelines that reinforce nonpartisanship and objectivity, and despite the harassment and hostility that many reporters confront online (Davis Kempton, S., & Connolly-Ahern, 2022), journalists have long used the platform to disclose personal information, share opinions about news stories, and socialize with others (Hermida, 2018; Lasorsa et al., 2012; Lee et al., 2016), making Twitter an ideal venue in which to see how journalists contribute to shaping narratives about AI.

**The Current Paper**

Taken together, we attempt to understand the relationship between journalists' language patterns and emotions, before and after the launch of ChatGPT. This work is timely and important because it is presently unclear how tastemakers of technology thought and felt about such a consequential tool at the time of its rise into public consciousness and mainstream use.

The case of journalists presents a previously underappreciated, but vital, dimension of this taste-making role in society. Thus, examining the reactions of journalists offers a window into their influence as key sense-makers for how people come to perceive and eventually use emerging technologies such as AI (see Brennen et al., 2022). We applied various computational and natural language processing techniques to consider how emotions were reflected and revealed over time during this critical moment in technological and generative AI history. Against this backdrop, we propose the following overarching research question:

*RQ:* What is the relationship between journalists' reaction to advances in artificial intelligence and the manifestation of emotion?

## Method

### Data Collection

To evaluate how journalists focused on emotion in their public disclosures before and after the ChatGPT launch, we gathered Twitter handles from a collection of journalists among 18 major news outlets in the U.S. ($n = 4,071$ unique accounts).[1] We used the Twitter lists from each source to access journalist Twitter handles.[2] Using the Academic Twitter API (Barrie et al., 2022), we extracted all Tweets two months before ($n = 534,757$ Tweets) and two months after ($n = 424,623$ Tweets) the launch of ChatGPT from each handle (November 30, 2022). Our database contained a total of 959,380 Tweets.

### Automated Text Analysis

We used an automated text analysis tool, Linguistic Inquiry and Word Count (LIWC), to

---

[1] The 18 sources *were The New York Times, The Washington Post, The Wall Street Journal, The Associated Press, USA Today*, NBC News, CBS News, ABC News, NPR News, PBS News, CNN, Fox News, MSNBC, *The Atlantic, The New Yorker, Wired, Time*, and Buzzfeed.

[2] For example, the URL for *The New York Times* journalists Twitter List is https://twitter.com/i/lists/54340435. All Twitter handles in this list were extracted, and this process was repeated for each outlet.

examine the rate of emotion across Tweets written before and after the ChatGPT launch

(Pennebaker et al., 2022). LIWC is a gold-standard text analysis program that has been used

extensively in the social sciences to evaluate social and psychological dynamics like emotion

(Boyd & Schwartz, 2021; Pennebaker, 2011; Tausczik & Pennebaker, 2010). The program

counts words as percentage of the total word count per text, identifying if words are found in its

internal dictionary of social (e.g., words related to friends), psychological (e.g., words related to

cognition, emotion), and part of speech categories (e.g., articles, prepositions). For example, the

statement "I believe AI will be great for the world" contains 9 words, and LIWC identifies the

following categories, including but not limited to self-references (*I*; 11.11% of the total word

count) and positive emotion terms (*great*; 11.11%). All language dimensions were drawn from

the standard LIWC-22 dictionary, and each Tweet received a score (e.g., the percentage of each

Tweet containing a verbal dimension of interest) across all measures of tone and emotion (see

Table 1). Given the size and scale of the data, we did not pre-process the texts and they were therefore analyzed as they appeared online.

**Measures**

We took a layered approach to evaluate the relationship between Time (before ChatGPT

vs. after ChatGPT) and emotion by evaluating different aspects of emotion: (1) discrete positive and negative emotions, and (2) positive and negative sentiment (tone). As prior work suggests, there are six discrete emotions (sadness, happiness, fear, anger, surprise, disgust) (Barrett, 2017),

and LIWC has two separate categories to approximate discrete positive emotions (e.g., words

such as *amaze*, *awesome*, and *excellent*) and discrete negative emotions (e.g., words such as

*agitate*, *suffer*, and *terrify*). Sentiment, or tone, describes the general positive or negative feeling that a text may elicit (Preoţiuc-Pietro et al., 2016). This dimension is linked to, but still independent from, emotion. For example, the word *birthday* is a positively-valenced term and elicits a positive tone, but the word is not a discrete emotion. We therefore evaluated two discrete emotion categories (i.e., emo_pos and emo_neg in LIWC) and two sentiment or tone categories (i.e., tone_pos and tone_neg in LIWC).

**Analytic Plan**

Using the *lme4* and *lmerTest* packages in R (Bates et al., 2015; Kuznetsova et al., 2020), we computed linear mixed models with a random intercept for Tweet writer to control for data non-independence. Four models were computed (one for each dependent variable), predicting emotion or sentiment from *Time* (before ChatGPT launch vs. after ChatGPT launch).

<div align="center">

**Results**

</div>

Estimated marginal means and effect sizes for each model are represented in Table 1. Overall trends in emotion and tone over time are provided in Figure 1 for illustration purposes and should be interpreted with caution due to the constrained y-axes.

**Positive and Negative Emotion**

Writers focused on more positive emotion after the ChatGPT launch compared to before the ChatGPT launch ($t = 12.64$, $p < .001$). The relationship between time and negative emotion was not statistically significant ($t = 0.65$, $p = .519$).

**Positive and Negative Tone**

Writers had a more positive tone after the ChatGPT launch compared to before the ChatGPT launch ($t = 14.01$, $p < .001$). Writers also had a less negative tone after the ChatGPT launch compared to before the ChatGPT launch ($t = -6.60$, $p < .001$). Altogether, writers were more positive and less negative after the ChatGPT launch versus before the ChatGPT launch.

It is important to note that the positive emotion, negative emotion, and positive tone effects were maintained after accounting for the overall daily happiness rating of Twitter using the Hedonometer (Dodds et al., 2011, 2015). After controlling for overall Twitter happiness, the relationship between time and negative emotion was marginally significant where Tweets written before the launch of ChatGPT were more negative than Tweets written after the launch of ChatGPT ($B = 0.01$, $SE = 0.008$, $t = 1.65$, $p = .098$).

**Alternative Explanations**

One possible explanation for the prior effects is that journalists were reporting the general mood or interest level of the public, not their own mood or interest level as they learned about ChatGPT. We therefore evaluated the prior relationships between time and emotion after accounting for Google Trends data. Google Trends provides a metric of search term popularity by day, and we considered the popularity of the search term "chatgpt" in the US over time.

After accounting for Google Trends data as a fixed effect in our prior linear mixed model calculations, writers still focused on more positive emotion after the ChatGPT launch compared to before the ChatGPT launch ($B = 0.127$, $SE = 0.011$, $t = 11.30$, $p < .001$). Writers also focused on more negative emotion after the ChatGPT launch compared to before the ChatGPT launch after accounting for Google Trends data ($B = 0.056$, $SE = 0.014$, $t = 4.16$, $p < .001$); recall, this pattern was not statistically significant in our prior model. Finally, consistent with our prior results and compared to before the ChatGPT launch, writers had a more positive tone ($B = 0.309$, $SE = 0.021$, $t = 14.98$, $p < .001$) but a less negative tone after the ChatGPT launch ($B = -0.067$, $SE = 0.017$, $t = -4.05$, $p < .001$), upon controlling for Google Trends data.

A second alternative explanation for these effects is that the general increase in positivity and decrease in negativity over time is the result of context effects, namely that the holiday season was within the timeframe of interest. Perhaps people were generally more positive in December and January (compared to October and November) because of holidays, not ChatGPT. LIWC's *religion* category contains holiday-specific words (e.g., *Christmas*, *church*, *faith*, *temple*) and therefore provides a useful proxy for holiday words. All relationships listed in Table 1 were maintained after controlling for holiday words via the religion index.

## Discussion

Disruptive technologies such as artificial intelligence give rise to competing narratives around their benefits and risks in society. The introduction of generative AI in the well-publicized form of ChatGPT in late 2022 is no exception, with contrasting discourses of hope and fear surrounding its implications for business, creative industries, education, and other sectors. Journalists, known for their ability to shape public imagination by offering the "first draft of history" on many noteworthy issues and events, perform an important role as tastemakers in the ongoing narratives around emerging technologies (Brennen et al., 2022). Our research therefore offers at least three main contributions: First, it adds to the study of technology and society by providing a first-of-its-kind large-scale empirical investigation into the emotional reactions of journalists on Twitter/X during a critical juncture of potential technological disruption. Second, it responds to the call for an "emotional turn" in journalism studies: the need to empirically and theoretically understand emotions and how they interact with norms, values, and practices in journalism (Wahl-Jorgensen & Pantti, 2021). Third, by conducting an extensive analysis of language patterns on social media, we also offer a methodological approach for others to follow in attempting to surface how communities of influential tastemakers react to historical events such as technological disruptions. Ultimately, it matters to study these prominent reactions because they contribute to shaping public narratives about technologies and their impact—whether pro or con, hopeful or fearful. In turn, the stories that people tell themselves about technology influence not only how emergent technologies are understood in the present, but also how they are imagined for the future, with corresponding consequences for individual and collective decisions made around matters such as acceptance, adoption, and regulation (cf. Brennen et al., 2022).

This study thus time-stamps a key moment in (technology) history, and it provides, among other things, a record of community reactions toward this critical incident in journalism (cf. Tandoc et al., 2020). Our findings document that journalists met the development of ChatGPT with positivity, as evidenced by the significantly positive tone and emotion before and after the chatbot's public launch in November 2022. This positivity, however, is rather noteworthy because it was never a given. In fact, it may be rather surprising given the recent history of disruptive technologies in journalism, where a succession of innovations—from the internet to social media to the smartphone—have largely undercut the traditional business models for journalism, leading to the layoffs of thousands of reporters and editors, and at the same time have amplified the "digital demands" on journalists, forcing those still in the profession to do more and more with fewer and fewer resources (e.g., see Bélair-Gagnon et al., 2022, 2024; Lewis & Molyneux, 2018). As a result of the growing anxiety, stress, and burnout that many journalists report, many are simply calling it quits (Mathews et al., 2023). So, for journalists to express such positive emotion about the introduction of generative AI, a technology that some observers initially feared could replace human writers, amplifying the challenges facing an already troubled profession like journalism, might offer a surprising twist. Perhaps it says something about the number of journalists who might have seen in ChatGPT a tool that could take on such of their grunt work, providing a respite from the incessant "hamster wheel" of expectations surrounding digital publishing (Usher, 2016).

Indeed, if we look historically, people's reactions to technologies believed to be "labor-saving"—such as semi-automatic technologies like type-cast printing and the loom—were initially positive in some cases, particularly when such tools improved industries and elevated standards of living (Carlopio, 1988). The positive emotions we document in this study may also signal an initial inclination toward the hopeful imaginaries associated with artificial intelligence, both AI generally and generative AI particularly, which relates to how such technologies can be used for such labor-saving functions—making work more efficient, freeing journalists from manual tasks in reporting, and augmenting economic and social capacity broadly (Cave & Dihal, 2019; Hautala & Heino, 2023). These positive emotions also run parallel to how journalists initially viewed the internet in the early 2000s (Chadha & Wells, 2016), even if those feelings may have changed as digitally driven burnout increased (see examples in Bélair-Gagnon et al., 2024). The positive emotions evident among journalists about generative AI may also be indicative of hype cycles, as well as their affective manifestations, that tend to come with the introduction of new technologies (Bourne, 2024).

It must be acknowledged that as new and disruptive technologies become fully integrated into workflows, they may exacerbate existing unjust economic, social, and cultural conditions, leading to moral panics often associated with new technology (Carlopio, 1988). Historically, the mismanagement of such technologies has led to negative perceptions and even prompted workers to destroy the machines and tools they deemed threatening (Carlopio, 1988). In the recent past, the emergence of new technologies have contributed to increased expectations for journalists to possess the necessary skills to use these technologies as well as obligations to use them (Bélair-Gagnon et al., 2022; Lewis & Molyneux, 2018), but the real, negative effects of new technologies on the bases of these expectations is anticipated to be felt belatedly rather than immediately, much like the negative impacts of news digitalization. In other words, the negative emotions around generative AI may come *later* as the technology is widely used in newsrooms, which warrants a further examination of emotions of journalists down the line.

With these considerations in mind, it is also worth noting that our computational work offered major empirical advantages compared to other projects that might be interested in understanding the opinions of journalists in response to technological innovation. We used natural language from nearly 1 million social media posts to identify how thousands of journalists thought and felt about a technology in the moment that it was introduced. This metaphorical microscope, in terms of scale (e.g., the number of people and posts in the analyses) and scope (e.g., the longitudinal nature of the work), is often difficult to capture for most studies that tend to be cross-sectional in nature or contain only a small number of journalists as participants. Our work has taken advantage of such computational social scientific approaches to understand psychological information about journalists through their own words—specifically, how they considered, made sense of, and felt about a new technology with uncertain prospects for their own future and profession. We advocate for scholars (and newsrooms) interested in understanding journalists' opinions on technology to consider using such natural language processing techniques in their own work as well, as they offer a unique window into opinions that are immediate, unadulterated, and psychologically rich.

## Conclusion

Our research empirically documents journalists' responses to new technologies immediately following their introduction, drawing parallels to reactions to past technologies deemed revolutionary. Journalists, as tastemakers and sense-makers for society (Singer, 1998), can mobilize their emotions, and the public display of these emotions can construct myths and narratives that influence societal understanding of technology's import and impact (Wahl-Jorgensen & Pantti, 2021). We acknowledge that Twitter (now X) is only one of many spaces where journalists might express their reactions to a critical incident such as this one. However, our study captures a snapshot of the broader range of platforms and venues where journalists' emotions are displayed. Further, we are limited to correlational and not causal claims in this work due to the nature of the field study that was conducted. Despite this, Twitter's role in public discourse is well-established in scholarly work, and it remains a preferred platform for journalists to break news and share personal opinions. Future research could explore how this phenomenon unfolds across other platforms, such as Facebook, to provide more comprehensive insights into journalists' reactions as a tastemaking community and the integration of these reactions into the news narratives that unfold in the weeks, months, and even years thereafter.

We were also limited to data from journalists associated with large and mainstream media entities. This was purposeful given our tastemaker and sense-maker framing, and it reflects the reality that journalists at major national news organizations tend to be mimicked by journalists at regional and local media outlets (e.g., see the concept of intermedia agenda-setting; McCombs & Valenzuela, 2020). However, future work might use journalists in smaller markets or at more niche publications to identify how our results compare. Finally, we collected these data shortly before the Twitter API became more expensive to use to conduct academic research, which, unfortunately, may complicate future work of this kind. Access to social media data is important for academe and industry alike (Rathje, 2024)—a conversation we believe is just in its infancy and deserves attention as collectively we seek to address some of society's most pressing questions about human emotion, perception, and behavior.

**Table 1**

*Estimated Marginal Means Across Factors*

| Emotion dimension | Example | Before ChatGPT Launch | | After ChatGPT Launch | | | | | |
|---|---|---|---|---|---|---|---|---|---|
| | | *M* | *SE* | *M* | *SE* | *t* | *p* | *R²m* | *R₂c* |
| Positive emotion (%) | *good*, *love*, *happy* | 0.91 | 0.01 | 0.98 | 0.01 | -12.64 | < .001 | 1.66E-04 | 0.052 |
| Negative emotion (%) | *bad*, *hate*, *hurt* | 2.70 | 0.02 | 2.70 | 0.02 | 0.65 | .519 | 4.19E-07 | 0.085 |
| Positive tone (%) | *accomplish*, *improve* | 3.48 | 0.03 | 3.64 | 0.03 | -14.01 | < .001 | 1.96E-04 | 0.094 |
| Negative tone (%) | *fright*, *plague* | 3.75 | 0.02 | 3.69 | 0.02 | 6.60 | < .001 | 4.46E-05 | 0.068 |

**Figure 1**

*Descriptive Trends in Emotion Over Time*

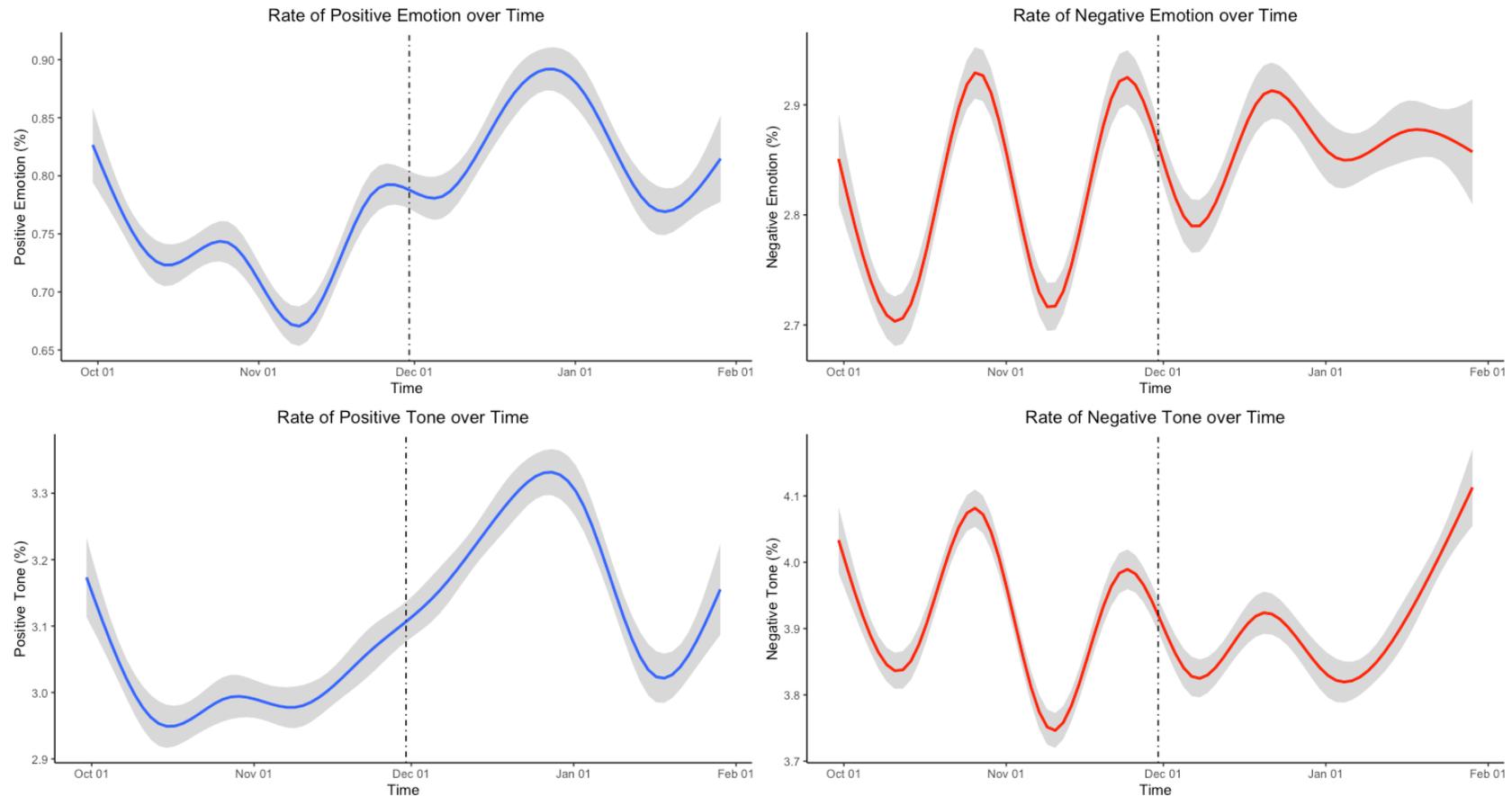

*Note.* Regression lines were estimated using the localized (LOESS) regression approach. We decided to not start these figures at the origin to demonstrate the dynamic nature of emotion over time at a local level. We do not encourage readers to use these figures to draw systematic conclusions about the link between time and emotion, as statistical tests are best for this purpose. The vertical line in each figure represents the date of ChatGPT's launch (November 30, 2022).